\documentclass[useAMS,usenatbib]{aa}  

\usepackage{graphicx}
\usepackage{txfonts}
\usepackage{graphicx} 
\usepackage{times}
\usepackage{epstopdf}
\usepackage{amsfonts}
\usepackage{amsmath}
\usepackage{amsbsy}
\usepackage{bm}
\usepackage{subcaption}
\usepackage{url}
\usepackage{microtype}
\usepackage{rotating}
\usepackage{sidecap}
\usepackage{longtable}
\usepackage{lscape}
\usepackage{rotating}
 \usepackage{array,multirow,graphicx}
\usepackage[export]{adjustbox}
\usepackage{amssymb,xcolor}
\usepackage{colortbl}
\usepackage{booktabs}
\usepackage{url}
\usepackage{wrapfig}
\usepackage{dblfloatfix}

\usepackage[utf8]{inputenc}

\begin{document} 

\title{Milliarcsecond-scale jet reorientations in GRS~1915+105\\ pointing towards a warped accretion disc}

\titlerunning{Milliarcsecond-scale jet reorientations in GRS~1915+105}
   \author{
          L.I. de Regt,\inst{1,} \inst{2}
          P. Atri,\inst{1,} \inst{2}
          S.E. Motta,\inst{3,} \inst{4}
          R.P. Fender,\inst{4,} \inst{5}
          J.C.A. Miller-Jones,\inst{6}
          D.R.A. Williams-Baldwin\inst{7}
          }
    \authorrunning{de Regt et al. }
   \institute{ASTRON, Netherlands Institute for Radio Astronomy, Oude Hoogeveensedijk 4, 7991 PD Dwingeloo, The Netherlands\\
              \email{regt@astron.nl}
        \and
        Department of Astrophysics/IMAPP, Radboud University, P.O. Box 9010, 6500 GL Nijmegen, The Netherlands
        \and
        Istituto Nazionale di Astrofisica, Osservatorio Astronomico di Brera, via E.\,Bianchi 46, 23807 Merate (LC), Italy
        \and
        Astrophysics, Department of Physics, University of Oxford, Keble Road, Oxford OX1 3RH, United Kingdom
        \and
        Department of Astronomy, University of Cape Town, Private Bag X3, Rondebosch 7701, South Africa
        \and
        International Centre for Radio Astronomy Research, Curtin University, GPO Box U1987, Perth, WA 6845, Australia
        \and
        Jodrell Bank Centre for Astrophysics, Department of Physics \& Astronomy, University of Manchester, Manchester M13 9PL, United Kingdom
             }
   \date{Received xx yy, 2026; accepted abc xx, 2026}

 
  \abstract
   {}
    {The black hole X-ray binary GRS~1915+105 exhibited persistent X-ray activity for more than 26 years after its discovery before entering a prolonged low X-ray flux state in 2018, while remaining bright at radio wavelengths. Multi-wavelength observations suggest that this low X-ray state results from the emergence of a complex layer of absorbing material, whose exact nature remains unknown. Since the X-ray emission is obscured, radio jet observations provide an unobstructed view of the physical changes associated with this phase.
    Recently published Very Large Array (VLA) and East Asian VLBI Network (EAVN) observations revealed substantial changes in the jet position angle (PA). Here, we present higher-resolution VLBI observations to investigate the connection between the obscuration event and the jet behaviour close to the launch region.
    }
    {We observed GRS~1915+105 with the Very Long Baseline Array (VLBA) and European VLBI Network (EVN) during the early X-ray obscuration phase across four epochs spanning 2019--2021, sampling a diverse range of jet and accretion behaviours. In this paper, we focus on the milliarcsecond jet morphology and position angle changes across the four epochs, and contextualize these in relation to the reported X-ray behaviour.}
    {Many of the PAs measured post-2018 are significantly offset from GRS~1915+105's pre-obscuration jet axis, which has a weighted mean of $142^{\circ} \pm 7^{\circ}$ on milliarcsecond scales. We measure a maximum offset of $57^{\circ} \pm 8^{\circ}$ from this axis, the largest such offset reported for GRS~1915+105 to date. Our four epochs reveal a large variety of jet behaviour, ranging from a stable, compact source with a large PA offset, to a bright, flaring source with a similarly large offset, to an extended jet with a PA closer to the pre-obscuration axis, and finally to a non-detection.}
    {The lack of measurable PA changes within individual 8-hour epochs rules out rapid Lense-Thirring precession of the inner disc as the direct driver of the jet reorientation. Instead, the timescales involved, together with the system's geometry and the apparent connection to the X-ray obscuration event, favour a radiation-driven warped accretion disc, in a configuration that could naturally explain the persistent obscuration.}


   \maketitle
%
\section{Introduction}\label{Introduction}

Black hole X-ray binaries (BHXBs) are binary systems containing a stellar mass black hole (BH) that accretes matter from a low mass companion star. This accretion results in the formation of a disc around the BH that is heated through viscous processes and can emit in X-rays \citep{Shakura1973}.
For decades, such systems provided the primary observational evidence for stellar-mass BHs, until the first detections of BH-BH mergers by gravitational-wave observatories \citep{Abbott2016}, the discovery of an isolated stellar-mass BH through gravitational lensing \citep{Sahu2022}, and the identification of optically-detected BH candidates in wide orbits \citep{El_Badry_2022, El_Badry_2023}.

BHXBs are usually transient, varying on timescales much shorter than those of supermassive black holes (SMBHs) lying at the centre of most galaxies. BHXBs typically spend most of their time in quiescence, occasionally going into relatively short periods of outburst reaching luminosities of the order $L \sim 10^{39} \rm \, erg \, s^{-1}$ \citep[e.g.,][]{Chen1997}. These outbursts tend to show a cyclical transition between various accretion states that are accompanied by mechanical feedback in the form of winds and jets \citep[e.g.,][]{Fender2004, Fender2016, Munoz-Darias_2026}.

Relativistic jets launched by black holes power some of the most energetic phenomena in the Universe. For example, gamma-ray bursts (GRBs) can produce jets with observed luminosities up to the order of $\sim$$10^{54} \, \rm erg \, s^{-1}$ \citep{Abdo2009, Greiner2009, Frederiks2013}, jets from active galactic nuclei (AGN) can dominate the energy output of entire galaxies \citep{Begelman1984, Fabian2012},
and their jet-driven outflows can extend over megaparsec scales \citep{Oei2024}.
 The humanly-accessible timescales and repeated outbursts of BHXBs make them excellent laboratories for studying jet formation and evolution.

The states of an X-ray binary (XRB) outburst can span several orders of magnitude in luminosity in just weeks or months \citep{Remillard_2006}. 
The literature divides this pattern into several states based on the spectral slope of the X-ray emission \citep{Remillard_2006} and X-ray timing properties of the emission \citep{Belloni_2009}. The `hard state', dominated by optically thin, Comptonized emission peaking at $\sim$$100 \, \rm keV$, exhibits strongly variable X-ray emission and is accompanied by steady radio jets. As the accretion flow becomes more radiatively efficient and the inner part of the disc moves inward, the source typically passes through intermediate states in which the Lorentz factor of the outflow is thought to increase and discrete, highly relativistic, optically thin ejecta are launched \citep{Fender2004}. In the subsequent soft state, the emission is dominated by optically thick thermal radiation with characteristic photon energies of $\sim$$1-2 \, \rm keV$. At this point, the jets have subsided and the main mode of mechanical feedback is in the form of disc winds.

Quasi-simultaneous X-ray and radio observations of BHXBs have established an empirical framework that links accretion flows to jet production \citep{Fender2004}.
In the hard state, this coupling manifests as a well-established radio-X-ray correlation \citep{Gallo2003, Corbel2003, Crook-Mansour2026}, which extends to AGN once mass scaling is taken into account \citep{Merloni2003, Falcke2004a}. Contemporaneous X-ray observations can therefore provide insights into the relationship between the accretion of inflowing matter and the launching of these jets.

Attempts to spatially resolve jets in XRBs have focused primarily on intermediate states, where discrete, large-scale ejecta can propagate to parsec scales from the core \citep{Hjellming1998, Yang_2010, Russell_2019, Bright_2020}. In contrast, jets in the hard state are typically confined within $\leq 10 \, \rm AU$ of the black hole, and only a handful have been spatially resolved to date \citep{Dhawan2000, Stirling2001, Ribo2004, Russell2015, Tetarenko2021, Wood2024}, likely due to a combination of source proximity, jet persistence, and the availability of sufficiently sensitive monitoring campaigns. As a result, the structure and dynamics of hard-state jets remain poorly constrained, and it is unclear whether the small number of resolved systems is representative of the broader population.

High-angular-resolution imaging of resolved hard-state jets provides direct measurements of fundamental jet properties -- such as opening angle, collimation, and brightness temperature. These complement constraints from radio and infrared timing studies \citep{Kalamkar_2016, Malzac_2018, Tetarenko_2019, Tetarenko2021}, astrometry and core-shift measurements \citep{Miller-Jones2021, Prabu_2023}, and broadband spectral modelling and spectral-break studies \citep{Markoff_2001, Corbel_2002}.

GRS~1915+105 is the largest known low-mass XRB (LMXB) and one of the best studied BHXBs in our Galaxy. It was first detected as a bright X-ray emitter in 1992 by the WATCH all-sky X-ray monitor on board the GRANAT space observatory \citep{Castro-Tirado1992} and soon after became the first Galactic source detected with apparently superluminal radio ejecta \citep{Mirabel1994}. Trigonometric parallax measurements of the system place it at a distance of $9.4 \pm 1.4 \, \rm kpc$ with an inclination angle $i = 64^\circ \pm 4^\circ$ \citep{Reid2023} and hosting a stellar mass BH \citep{Greiner2001} with a mass of $12.4^{+2.0}_{-1.8} \, \rm M_{\odot}$ \citep{Reid_2014}.

For over 26 years, GRS~1915+105 exhibited persistent, intense X-ray activity, until it entered a sudden prolonged period of low flux in July 2018 ($\sim \rm MJD \, 58300$), followed by further X-ray dimming around the end of March 2019 ($\sim \rm MJD \, 58600$)  \citep{Negoro2018, Motta2019}. Initially, this was interpreted as the end of GRS~1915+105's 26-yr long outburst, but this hypothesis was refuted only days later on 14 May 2019 ($\rm MJD \, 58617$) with the emergence of renewed flaring activity across different wavelengths.

Multi-wavelength observations of the X-rays, radio and near-infrared (NIR) winds have since indicated that this state of low X-ray flux is the result of enhanced obscuration, drawing similarities to the highly obscured Seyfert-2 AGNs 
\citep{ Miller_2020, Balakrishnan_2021, Motta2021, Sanchez-Sierras2023}. 
Several obscuring mechanisms have been proposed, including a puffed-up slim disc \citep{Abramowicz1998} at high inclination blocking our view of the inner accretion flow \citep{Neilsen_2020}, failed disc winds enshrouding the central engine \citep{Miller_2020, Sanchez-Sierras2023}, and a warped outer disc occulting the inner flow \citep{Miller_2025}. However, no single scenario has yet been definitively established as dominant.

The presence of intrinsic absorption has been reported only for a handful of other XRBs. In 2015, the BHXB V404~Cyg showed clear signatures of heavy and variable intrinsic absorption during both outbursts monitored in the X-rays, which were both interpreted as the result of a puffed-up disc blocking the view to the inner accretion flow \citep{Motta2017}. Similar disc-induced obscuration has also been inferred in the microquasar SS~433 \citep{Fabrika_2004, Middleton_2021}.
In both these systems, obscuration was the consequence of erratic (V404~Cyg) or sustained (SS~433) super-Eddington accretion, which in both cases is associated with extreme activity of the radio jets 
\citep{Spencer_1979, Miller-Jones_2019}.

Despite the suppressed X-ray flux, GRS~1915+105 has remained highly active at radio wavelengths, as shown by regular monitoring with the Arcminute Microkelvin Imager Large Array (AMI-LA) and MeerKAT \citep{Motta2021,Motta_2025}. 
Very Large Array (VLA) images at 8.0 GHz (Sep 30+Oct 01 2023) and 5.5 GHz (May 25th 2024) have shown a surprising change in the jet orientation angle at arcsecond scales \citep{Rodriguez2025}.
Additionally, East Asian VLBI network (EAVN) observations at 6.7 GHz and 43 GHz from 2023 April find a large misalignment between the discrete ejecta detected at 6.7 GHz ($\sim$$8^\circ$) and the continuous jets seen at 43 GHz ($\sim$$152^\circ$; \cite{Jiang_2026}). 
Most recent multi-wavelength monitoring found GRS~1915+105 at its lowest recorded activity
level since discovery, with non-detections in radio (L-band; \citet{Motta2026atel}), X-ray \citep{Marino2026atel}, optical \citep{Ducoin2026atel}, and near-infrared \citep{CotiZelati2026atel}, potentially indicating the onset of a genuine transition to quiescence after more than three
decades of continuous activity.

In this paper, we present four epochs of Very Long Baseline Interferometry (VLBI) observations targeting GRS~1915+105 at X-band (8.4 GHz) and C band (4.9 GHz), taken with the Very Long Baseline Array (VLBA) and the European VLBI Network (EVN), respectively. Being the earliest and highest-resolution radio observations of GRS~1915+105 during the obscured phase, these data allow us to study the jet emission on milliarcsecond scales, closest to the launch region where the accretion–outflow coupling is strongest and environmental effects on jet propagation are minimized.
In Section \ref{Methods} we describe our data reduction and imaging procedure, in Section \ref{sec: results} we present the observational results, in Section \ref{sec: analysis} we detail the analysis and parameter estimation, and in Section \ref{discussion} we discuss our findings.
Finally, in Section \ref{Conclusions} we summarize our main results and outline our
conclusions.

\begin{figure*}
\centering
\includegraphics[width=0.95\textwidth]{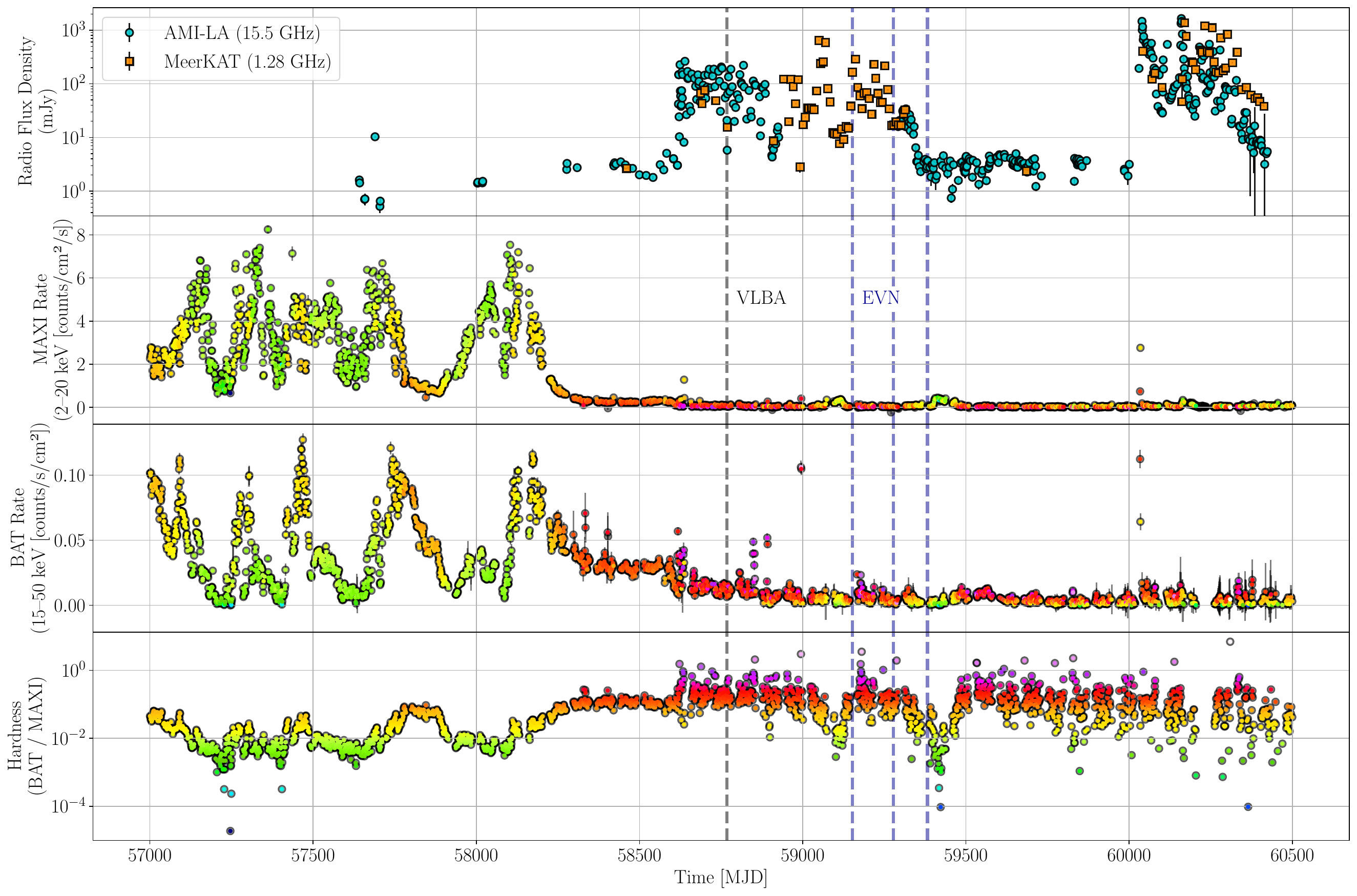}
\caption{AMI-LA and MeerKAT data, MAXI data, and BAT data, spanning approximately 10 years. The colour-coding indicates the X-ray spectral hardness, calculated as the ratio of the BAT to MAXI fluxes and shown numerically in the bottom panel.
The grey and purple dotted lines indicate the dates of our VLBA and EVN observations.
}
\label{fig:lightcurve} 
\end{figure*}

\begin{table*}
   \caption{Observation log of VLBA and EVN observations. The total time on source is denoted by $T_{\rm source}$. The central frequency is denoted by $\nu_c$.}
   \label{Observations}
   \centering
   \begin{tabular}{l c c c c c c l}
      \hline\hline
      Instrument & Proposal Code & Date & $T_{\rm source}$ & Time & $\nu_c$ & Bandwidth & Synthesized beam \\
             &      &      &  (minutes)   & (UTC) & (GHz) & (MHz)  & (mas $\times$ mas, $^\circ$ N to E)\\
      \hline
      VLBA$^a$ & BM500 & 2019-10-12 & 260.9 & 20:48 -- 03:48$^c$ & 8.4 & 256 & 1.03 $\times$ 2.39, 171.35 \\
      EVN$^b$  & EM144A & 2020-10-30 & 235.5 & 12:30 -- 20:30 & 4.9 & 256 & 1.39 $\times$ 3.50, 6.84 \\
      EVN$^b$  & EM144B & 2021-03-05 & 235.5 & 04:00 -- 12:00 & 4.9 & 256 & 2.49 $\times$ 3.11, 67.21 \\
      EVN$^b$  & EM144C & 2021-06-17 & 235.5 & 21:30 -- 05:30 & 4.9 & 256 & 1.49 $\times$ 2.83, 9.69 \\
      \hline
   \end{tabular}
   \tablefoot{
   \tablefoottext{a}{VLBA participating stations: Sc, Hn, Nl, Fd, La, Pt, Kp, Ov, Br, and Mk.}
   \tablefoottext{b}{EVN participating stations: Jb, Ef, Wb, Mc, O8, T6, Ur, Tr, Yb, Hh, Sv, Zl, Bd, Ir, and e-MERLIN stations Cm, Da, De, Kn, and Pi.}
   \tablefoottext{c}{Times denote on-source time and exclude the 30-min geodetic blocks at the start and end of the VLBA observation.}
   }
\label{tab: observations}
\end{table*}

\section{Observations and Calibration}
\label{Methods}

GRS~1915+105 was observed across four 8-hour epochs using VLBI: one at X-band (8.4\,GHz) with the VLBA (project code BM500, PI: S.E. Motta) and three at C-band (4.9\,GHz) with the EVN (project code EM144, PI: S.E. Motta). Figure \ref{fig:lightcurve} shows the timing of these observations overlaid on the radio flux density, X-ray flux and hardness evolution of GRS~1915+105 over the last $\sim$10 years. The X-ray fluxes are publicly available on the web pages of the Monitor of All-sky X-ray Image \citep[MAXI\footnote{\url{http://maxi.riken.jp/top/}};][]{Matsuoka_2009} and the \texttt{Swift}/BAT transient monitor \citep[BAT\footnote{\url{https://swift.gsfc.nasa.gov/results/transients/}};][]{Krimm_2013}. The observation log is presented in Table \ref{tab: observations}.

\subsection{VLBA}
\label{subsec: VLBA}
We observed GRS~1915+105 following the radio-brightening during its X-ray obscured state \citep{Motta2021} for 8 hours on October 12th 2019 with the VLBA. The observations were carried out in X-band (8.4\,GHz) with a recording rate of 2048\,Mbps, utilizing a total bandwidth of 256 MHz, divided into eight intermediate-frequency (IF) pairs of 32\,MHz each.
All ten VLBA stations participated: Saint Croix, Hancock, North Liberty, Fort Davis, Los Alamos, Pie Town, Kitt Peak, Owens Valley, Brewster, and Mauna Kea, although Los Alamos was flagged during calibration due to poor fringe fitting solutions. 

Of the total 8\,hours of observations, 260.9\,minutes were spent on target.
We observed J1800+3848 (7.3\,minutes) and J2253+1608 (2.9\,minutes) as fringe finders, J1905+0952 (86.6\,minutes) as a phase reference calibrator and J1913+0932 (8.5\,minutes) as a check source. The phase reference calibrator is at a 2.58$^\circ$ separation from the target. We alternated between 180\,second scans on the target and 60\,second scans on the phase calibrator, J1905+0952. The check source, J1913+0932, was observed for 60\,seconds approximately every 40\,minutes.
Furthermore, we used geodetic blocks \citep{Reid_2009} of $\sim$30\,minutes at the start and end of the observations to correct for excess tropospheric delays and clock errors. 

The data were correlated using the DiFX software correlator \citep{Deller_2007, Deller_2011}, and calibrated following the standard procedures within the Astronomical Image Processing System (\textsc{AIPS}, version 31DEC25; \citealt{Wells_1985, Greisen2003}). After applying standard external gain calibrations, we performed several iterations of self-calibration and imaging on the phase-reference calibrator to construct an optimal model. From this model, we derived time-varying phase, delay, and rate solutions, which were interpolated to the target source. The amplitude gain solutions obtained from the phase-reference source were also applied to the target to further improve the time-varying amplitude calibration.

\subsection{EVN}
GRS~1915+105 was observed for three epochs of 8 hours on October 30th 2020, March 5th 2021 and June 17th 2021 with the EVN. 
All these observations were performed in the C-band with a central frequency of 4.9\,GHz and eight IF pairs of 32\,MHz each. The participating EVN stations were Jodrell Bank (Lovell Telescope), Effelsberg, Medicina, Onsala, Toruń, Urumqi, Tian Ma, Yebes, Hartebeesthoek, Svetloe, Zelenchukskaya, Badary, and Irbene, together with the e-MERLIN stations Cambridge, Darnhall, Defford, Knockin, and Pickmere. The data were correlated with the SFXC software correlator at the Joint Institute for VLBI ERIC (JIVE).

The total on-source observation time was 235.5\,minutes per epoch.
Consistent with the VLBA observations, we used J1800+3848 (9.9\,minutes) and J2253+1608 (6.0\,minutes) as fringe finders, J1905+0952 (168.4\,minutes) as a phase reference calibrator and J1913+0932 (22.2\,minutes) as a check source. We alternated between 210\,second scans on the target and 120\,seconds scan on the phase calibrator. The check source was observed for 150\,seconds every fifth cycle.

The pointing centre during the EVN observations did not account for the source's proper motion. To correct for this, we applied a phase shift prior to fringe fitting using the AIPS task \texttt{CLCOR}. For the correction, we adopted the proper motion measured by \citet{Reid2023}: $\mu_\alpha = -3.14 \pm 0.03~\mathrm{mas\,yr^{-1}}, \quad \mu_\delta = -6.23 \pm 0.04~\mathrm{mas\,yr^{-1}}$. Calibration followed the standard procedures in \textsc{AIPS} similarly to the steps performed for the VLBA observation. 
To maximize sensitivity, we constructed the phase calibrator model from the combined calibrated visibilities of all three EVN observations before applying the model for phase referencing.

\section{Results} 
\label{sec: results}

\begin{figure*}[t]
\centering
\begin{subfigure}{0.325\textwidth}
    \includegraphics[width=\linewidth]{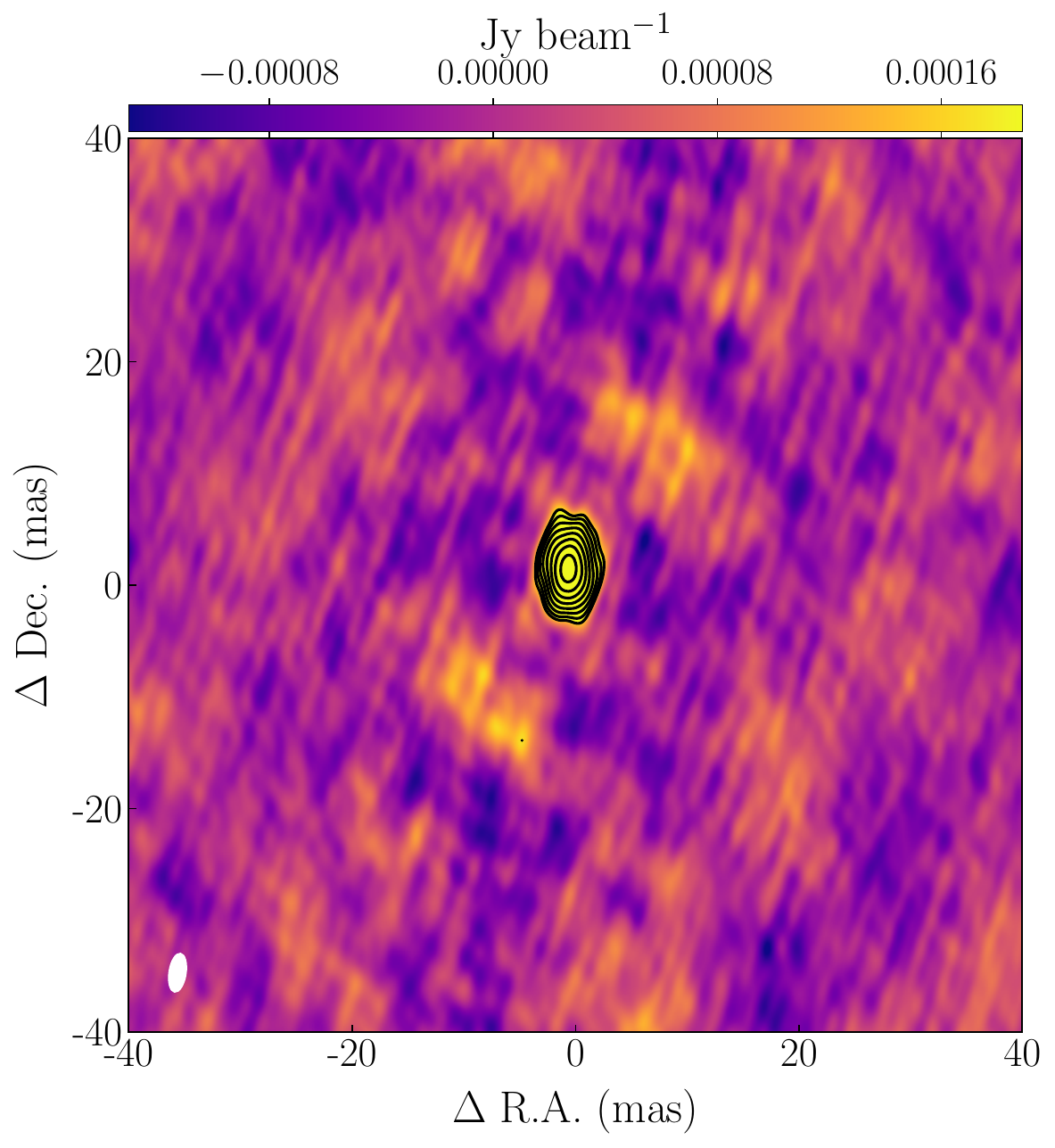}
    \caption{VLBA (12-Oct-2019 ; 8.4 GHz)}
\end{subfigure}
\begin{subfigure}{0.32\textwidth}
    \includegraphics[width=\linewidth]{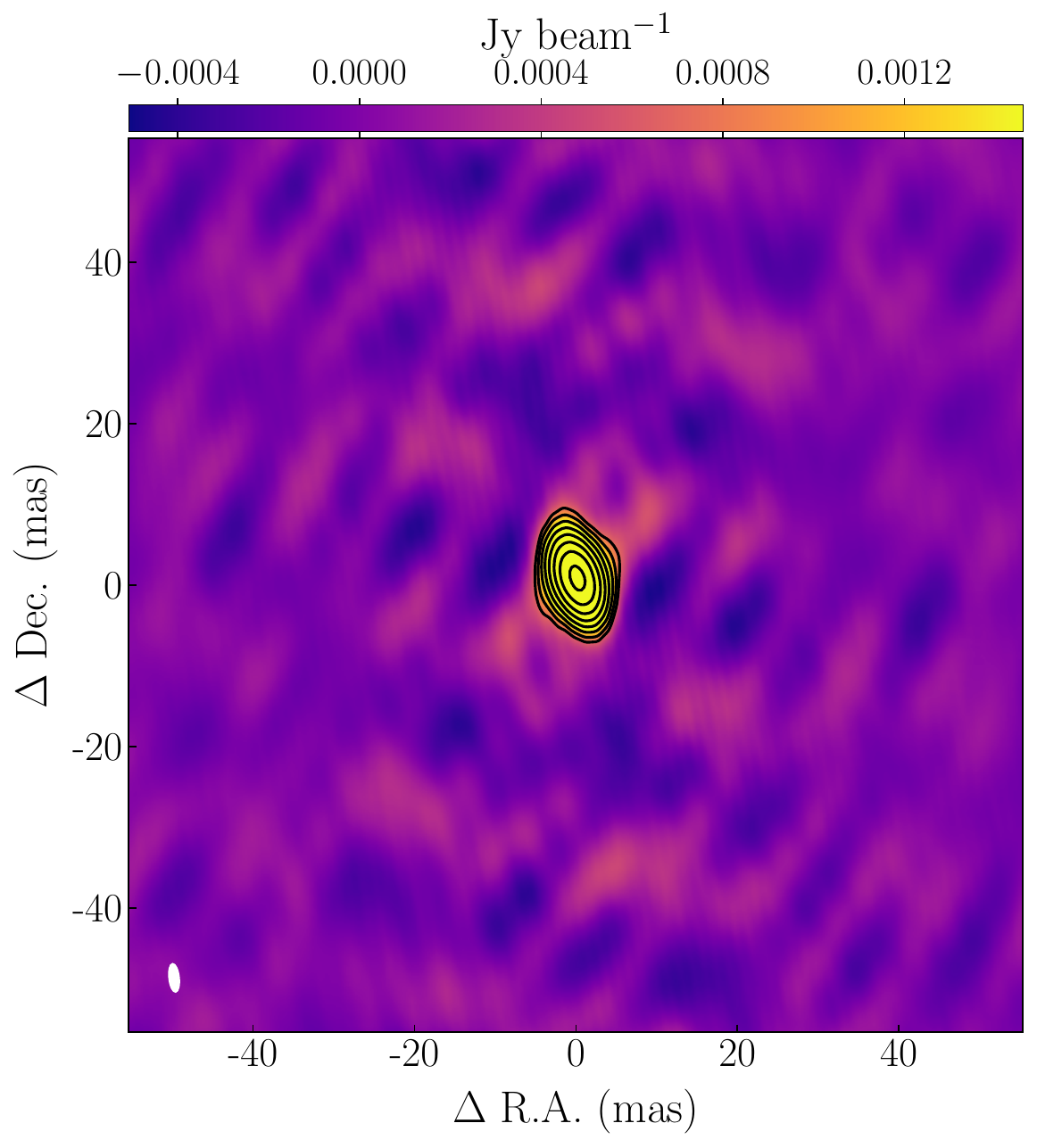}
    \caption{EVN E1 (30-Oct-2020 ; 4.9 GHz)}
\end{subfigure}
\begin{subfigure}{0.32\textwidth}
    \includegraphics[width=\linewidth]{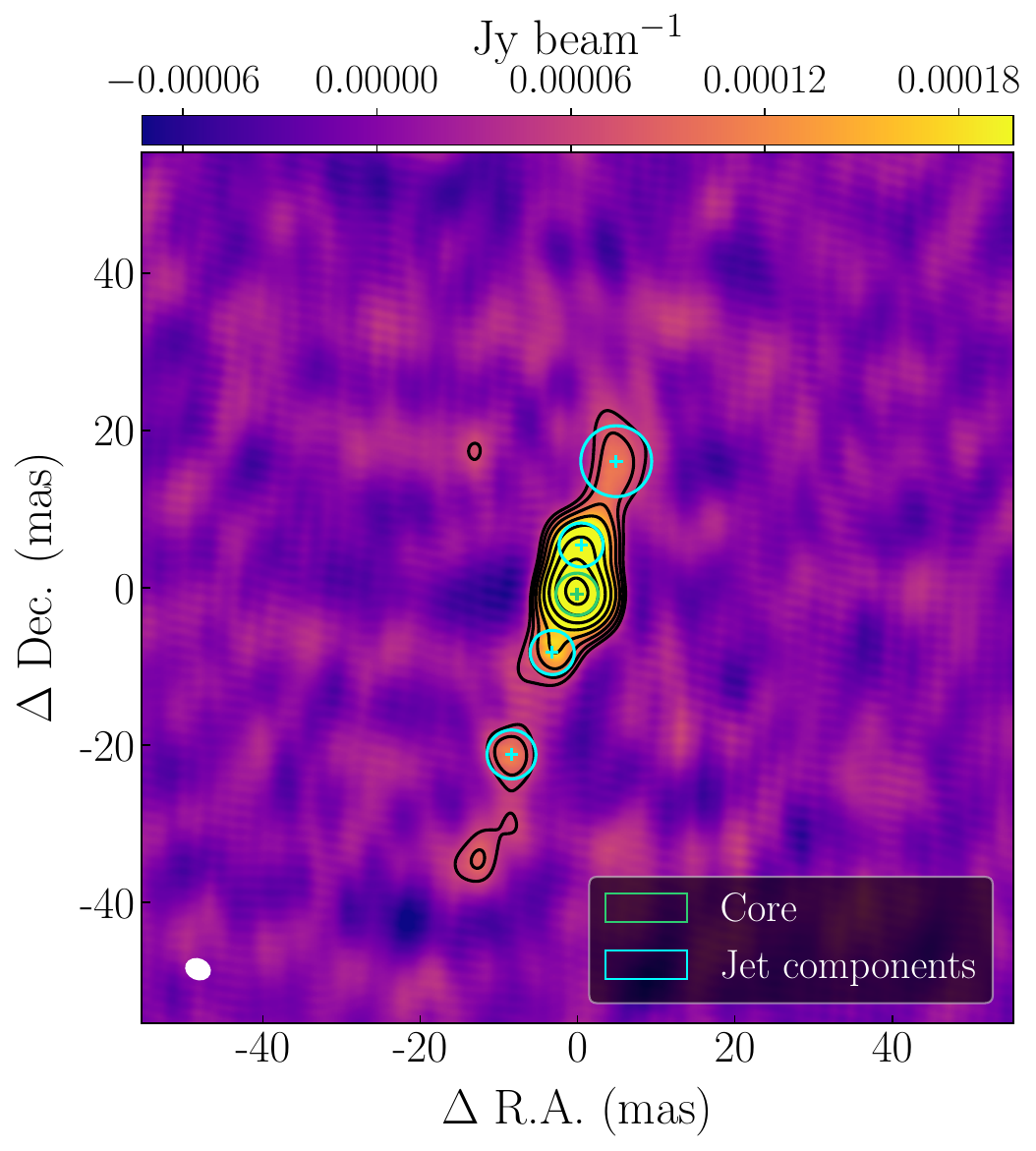}
    \caption{EVN E2 (05-Mar-2021 ; 4.9 GHz)}
\end{subfigure}
\caption{Natural weighting contour plots at the $3\sigma$ level of the three detections of GRS~1915+105. Panel (b) shows only the stable-flux interval 12:30–17:30 UT (see section \ref{sec: flaring}). The right ascension (R.A.) and declination (Dec.) offsets are relative to the best fit position.
Notice that the imaged field of the EVN epochs is larger than the size of the imaged VLBA field. 
EVN E2 (panel (c)) has the Gaussian model components overlaid onto the image.
} 
\label{fig:VLBA_threeplots}
\end{figure*}

\begin{table*}
\centering
\begin{tabular}{lccccccc}
\hline\hline
Epoch & $\theta_\text{maj}^\text{obs}$ & $\theta_\text{maj}^\text{deconv}$ & $S_\text{peak}$ & $S_\text{tot}$ & PA & $T_{B,\text{beam}}$ & $T_{B,\text{scat}}$ \\
      & (mas) & (mas) & (mJy\,beam$^{-1}$) & (mJy) & ($^\circ$) & (K) & (K) \\
\hline
VLBA   & $2.6$  & $1.7 \pm 0.2$ & $2.20 \pm 0.04$ & $6.23 \pm 0.62$ & $15.3 \pm 1.2$    & $\geq (6.2 \pm 1.0) \times 10^{7}$ & $\geq (3.7 \pm 0.6) \times 10^{7}$ \\
EVN E1 & $7.9$  & $5.6 \pm 0.6$ & $8.76 \pm 0.09$ & $90.3 \pm 9.0$  & $18.61 \pm 0.03$  & $\geq (6.8 \pm 1.0) \times 10^{8}$ & $\geq (1.5 \pm 0.2) \times 10^{8}$ \\
EVN E2 & $10.8$ & $9.2 \pm 0.3$ & $0.68 \pm 0.02$ & $8.12 \pm 0.81$ & $170.9 \pm 0.3$   & $\geq (1.3 \pm 0.1) \times 10^{7}$ & $\geq (5.8 \pm 0.6) \times 10^{6}$ \\
EVN E3 & $-$    & $-$           & $0.016 \pm 0.008$ & $< 0.04^{a}$    & $-$               & $-$ & $-$ \\
\hline
\end{tabular}
\caption{Gaussian fit parameters, peak flux density, and brightness temperature
lower limits for the VLBA and EVN observations, measured in \textsc{Difmap}.
$\theta_\text{maj}^\text{obs}$ and $\theta_\text{maj}^\text{deconv}$ represent the observed and scatter-broadening-corrected major-axis sizes of elliptical Gaussians fitted to the brightest central component. $S_\text{peak}$ is the peak flux density of the image, with error given by the image RMS noise; this reflects only the image-plane statistical noise and does not include the additional $\sim$10\% flux-scale calibration uncertainty adopted elsewhere.
$S_\text{tot}$ is the total integrated flux density of the fitted Gaussian, and PA the position angle of the fitted Gaussian. $T_{B,\text{beam}}$ and $T_{B,\text{scat}}$ are lower limits on
the brightness temperature computed using the array beam minor axis and the scattering size as upper limits on $\theta_\text{min}$, respectively. The statistical errors on $\theta_\text{maj}^\text{obs}$ are negligible compared to the assumed 10\% errors on the scatter broadening size and are therefore omitted.}
\tablefoot{
\tablefoottext{a}{$5\sigma$ upper limit on $S_\text{tot}$, based on the image RMS of $0.008$\,mJy\,beam$^{-1}$.}
}
\label{tab:source_params}
\end{table*}

The calibrated visibilities were imaged in \textsc{Difmap} \citep{Shepherd1994} using natural weighting to maximize sensitivity to extended emission. Component fitting was performed directly on the visibilities using the \texttt{MODELFIT} task, fitting circular or elliptical Gaussian components to the brightest emission regions. 

The results of the Gaussian fits to the calibrated data are detailed in Table~\ref{tab:source_params} and shown in Figure~\ref{fig:VLBA_threeplots} as contour plots with natural weighting.  We add 10\% systematic errors in quadrature with our statistical errors to account for inaccuracies of the absolute amplitude calibration. All detections show extended emission, indicating that the source is resolved on milliarcsecond scales.

\subsection{VLBA}

The detection in the VLBA epoch is best described by a single elliptical Gaussian component; the major and minor axes differ at the $\sim$$4\sigma$ level ($2.55\times 2.55$ mas circular vs $2.55 \times 1.84$ mas elliptical), though the improvement in reduced $\chi^2$ is marginal ($\Delta\chi^2_\nu \approx 10^{-4}$).
The major axis is oriented at $15.3^\circ \pm 1.2^\circ$ and is $23.9^{\circ}$ offset from the beam orientation ($171.4^{\circ}$), confirming that the elongation reflects intrinsic source structure rather than beam smearing. The deconvolved major axis of $1.7 \pm 0.2 \, \rm mas$ implies the source is resolved at 8.4\,GHz on milliarcsecond scales. Despite this resolved structure, the fitted Gaussian centroid provides a robust position estimate for the core: RA $=19^\mathrm{h}15^\mathrm{m}11.5448325^\mathrm{s} \pm 0.102\,\mathrm{mas}$, Dec $=+10^\circ56'44.626020'' \pm 0.121\,\mathrm{mas}$.

\subsection{Flaring activity during EVN epoch 1}
\label{sec: flaring}

\begin{figure*}
\centering
\includegraphics[width=0.95\textwidth]{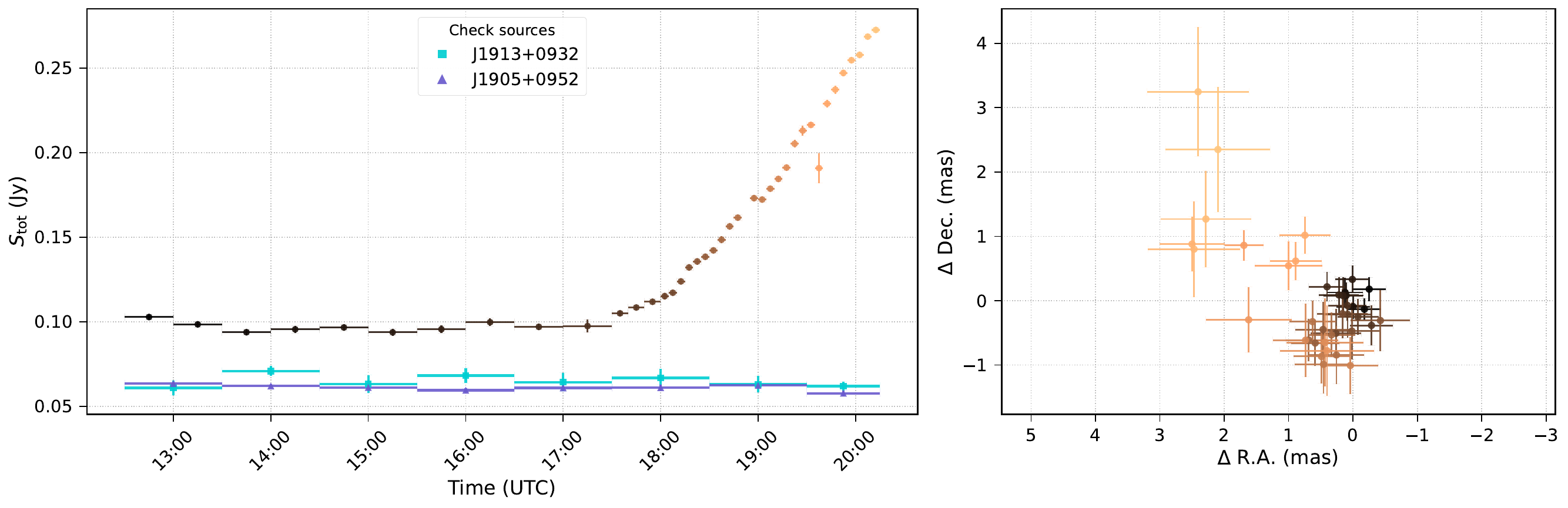}
\caption{Integrated source flux density evolution (left) and the on-sky best fit positions (right) of the highly varying radio emission during EVN E1 (EM144A). The observation has been split into time bins in a manner to allow for a maximum of $\sim 10$\% change in amplitudes. The integrated flux density of GRS~1915+105 is compared with the phase calibrator (J1905+0952) and check source (J1913+0932) to rule out calibration errors. Data points are colour-coded by time, allowing corresponding flux density and position measurements to be identified across the two panels. The nullpoint of the position plot is taken to be the mean of the stable flux density positions (before 17:30 UTC).
}
\label{fig:EM144A} 
\end{figure*}

During our first EVN epoch (EVN E1 hereafter), the source showed a dramatic increase in flux density, rising by over $150~\mathrm{mJy}$ within 3 hours (Fig.~\ref{fig:EM144A}).  
Such rapid variability suggests that GRS~1915+105 was observed during a flaring event, typically associated with the ejection of discrete jets.

Given the high flux density of the source, we divided the observation into 5-minute time bins, ensuring that the amplitude variation between consecutive bins remains within $\sim$10\%.
We identified a stable-flux region (12:30–17:30 UTC) and a rising-flux region (17:30–20:30 UTC). This allows for imaging the stable part of the observation directly using conventional imaging techniques (Fig. \ref{fig:VLBA_threeplots}(b)). During this stable interval, we imaged the source in 30-minute segments and found that its position, flux density, and position angle remained constant within uncertainties.

The centroid positions from individual time bins were fitted using \texttt{JMFIT} (Fig.~\ref{fig:EM144A}). The integrated source flux density measurements are measured following self-calibration on 10 minute intervals, to optimize the amplitude calibration. The position fits are applied without self-calibration to retain absolute position accuracy.

Figure~\ref{fig:EM144A} (left) presents the integrated source flux density evolution of GRS~1915+105 within the 8-hour observation and compares this with the check source and phase calibrator flux density over time. The roughly constant flux density for both of these sources rules out amplitude calibration errors as the cause of the flux density increase in GRS~1915+105.
The comparatively low flux density in the range 19:35-19:40 UTC is due to particularly poor $uv$-coverage within this time bin. The overall shape of the flux density evolution roughly follows a hockey-stick function. However, whether this reflects a linear flux density increase from a single component or a superposition of multiple independently varying components remains an open question, which we will pursue in a future paper.

Figure~\ref{fig:EM144A} (right) shows the best fit positions of the fitted components in each time bin.
Systematic uncertainties were added following \cite{Pradel2006} and \citet{Reid_2017}.  
We find a systematic shift in position along the axis of elongation, coinciding in time with the rise in flux density. This positional shift along the jet axis is consistent with measurable jet motion coincident with the onset of the radio flaring.
No comparable shift is seen in the position of the check source over the same time bins, indicating that the observed motion is not an artifact of short timescale ionospheric variations or residual phase-referencing errors.
However, splitting into smaller time bins severely limits the $uv$-coverage per data point, which greatly reduces our astrometric accuracy and could still, in principle, increase the risk of short timescale ionospheric variations affecting our positions.

A more detailed treatment of the implied jet velocity, and its relationship to the position angle behaviour discussed in Section~\ref{subsec:jet_changes}, will be the subject of a forthcoming study incorporating additional EVN monitoring of GRS~1915+105, and will allow a direct test of the recently proposed jet paradigm by \citet{Fender2025}, which distinguishes between precessing, slow jets and fixed, fast jets, that arise from the inner disc locking onto the black hole spin-axis. 

\subsection{EVN epoch 2}
EVN epoch 2 (hereby EVN E2) reveals an elongated core component with additional emission extending along the same orientation angle. We model the source with a total of five circular Gaussian components (overlaid on Fig. \ref{fig:VLBA_threeplots}c). All components are aligned with the brightest central emission, as well as the jet orientation angle observed with the VLA in 2023 \citep{Rodriguez2025}. To rule out artefacts resulting from amplitude calibration errors we carried out tests including removing individual antennas, applying $uv$-tapers, varying the robustness parameter, and attempting self-calibration on several solution intervals.

\subsection{EVN epoch 3}
The third epoch of EVN observations (EVN E3) resulted in a non-detection and, therefore, is not shown in Figure \ref{fig:VLBA_threeplots}. The calibrated data were also imaged using only the eMERLIN dishes, Westerbork, Jodrell Bank and Effelsberg to check for extended emission that might otherwise be resolved out by the full array, and to search for possible large-scale ejecta. No large-scale ejecta were
detected. This non-detection places a $5\sigma$ upper limit of $\leq0.04\,\rm mJy$ on the source's flux density, indicating the source had faded by a factor of $\gtrsim220$ relative to its EVN~E1 peak flux density eight months earlier. AMI-LA observations taken one day before and one day after EVN~E3 measured a flux density of $\sim0.15$\,mJy\,beam$^{-1}$ at
$15.5$\,GHz. Given AMI-LA's much larger, tens-of-arcsecond-scale beam compared to our milliarcsecond-resolution VLBI observations, this is not necessarily inconsistent with our non-detection. Any diffuse or extended emission contributing to the AMI-LA flux density would be resolved out by the long VLBI baselines, and would not require a compact source above our detection threshold.


\section{Analysis}
\label{sec: analysis}

\subsection{Source morphology and extension}
GRS~1915+105 lies close to the Galactic plane, where the line of sight intercepts a large column of ionized gas. This produces significant angular broadening and is partially responsible for the observed extension visible in our images. The \texttt{NE2025} Galactic electron density model \citep{Ocker_2026} predicts a scattering size of $1.4$~mas at $5\,\rm GHz$ at the location of GRS~1915+105. \citet{Dhawan2000} found a scattering size of $\sim 1.9$~mas at $8.4~\mathrm{GHz}$ and $\sim 5.6$~mas at $4.9~\mathrm{GHz}$ for GRS~1915+105. We prefer the direct, source-specific scatter-broadening measurements of \citet{Dhawan2000}
over the line-of-sight-averaged prediction from \texttt{NE2025} for this analysis.

The \citet{Dhawan2000} values agree with the observed extension of the minor axes fit by \texttt{JMFIT} in all epochs. We assume isotropic scattering and, consequently, account for the angular broadening by assuming a circular Gaussian such that $\theta_\mathrm{deconv}^2 = \theta_\mathrm{obs}^2 - \theta_\mathrm{scat}^2$. We adopt a conservative 10\% uncertainty on the scattering size. After decoupling the measured source sizes from the expected scatter-broadening, the
intrinsic source sizes still show significant elongation (Table~\ref{tab:source_params}; for beam sizes, see Table~\ref{tab: observations}),
suggesting the presence of continuous jet production. At the source distance of $9.4\pm1.4$\,kpc \citep{Reid2023}, the deconvolved major-axis sizes correspond to projected extents of $16.0\pm3.0$\,AU (VLBA, $8.4$\,GHz), $52.6\pm9.7$\,AU (EVN~E1, $4.9$\,GHz), and $86.5\pm13.2$\,AU (EVN~E2, $4.9$\,GHz). These are broadly consistent with the compact continuous jet sizes of a few to $\sim100$\,AU reported by \citet{Dhawan2000}.

We estimate the brightness temperature based on the measured source size and flux density according to the Rayleigh-Jeans formalism \citep{Readhead_1994, Lee_2008}
\begin{equation}
    T_b \approx 1.22 \times 10^{12} \frac{S_\nu [{\rm Jy}]}{\nu^2 [{\rm GHz}] \theta_{maj} [{\rm mas}] \theta_{min} [{\rm mas}] }  \rm{K}
\end{equation}
The minor axes are limited by the synthesized beam and scatter broadening, yielding lower limits on the true brightness temperatures.
The lower limits on the brightness temperatures, ranging from $\sim6\times10^6 \, \rm K$ to $\sim7\times10^8$\,K across epochs (Table~\ref{tab:source_params}), are consistent with non-thermal, synchrotron emission, as expected for either a compact or extended radio jet.

\subsection{Jet inclination angles}
In addition to the changes in X-ray flux and jet position angle discussed above, if these position angle variations reflect precession of the jet axis, a corresponding change in the inclination angle would also be expected. Prior to obscuration, the inclination angle was consistent for compact and discrete ejecta over 25 years, with a best estimate of $64 \pm 4^\circ$ \citep{Reid2023}. Recent measurements of discrete jet ejecta during the obscured phase suggest a nearly edge-on geometry, $i=87 \pm 3^\circ$ \citep{Rodriguez2025} and $i=89 \pm 1^\circ$ \citep{Jiang_2026}. This motivates constraining $i$ using our data.

\begin{figure}
\centering
\includegraphics[width=0.9\columnwidth]{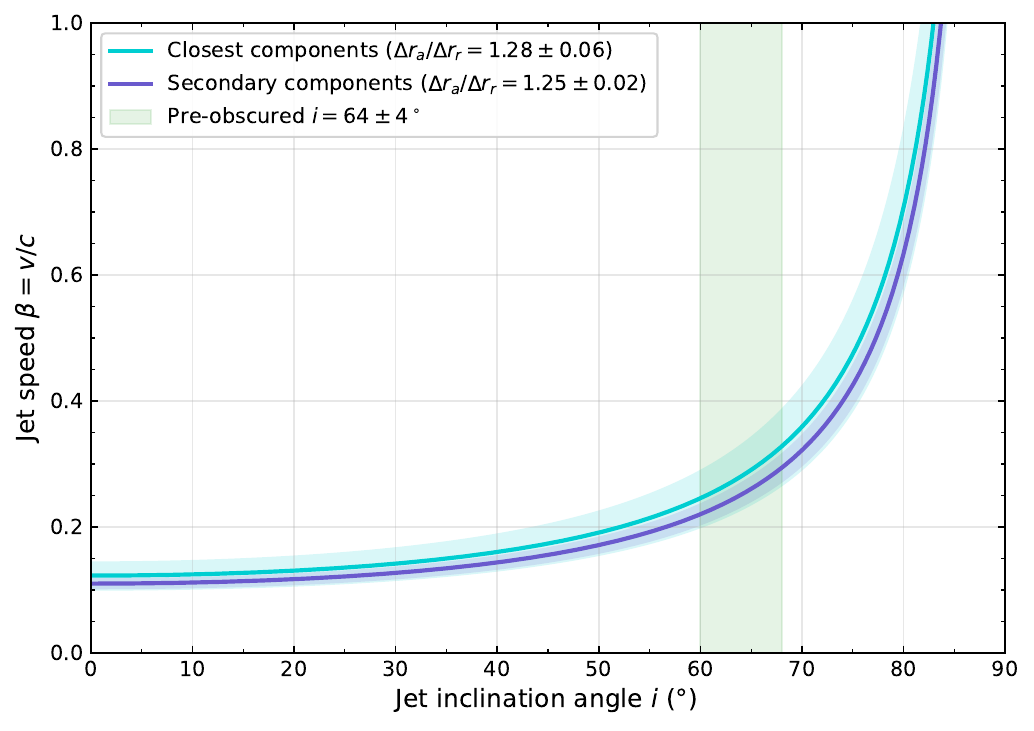}
\caption{\footnotesize
Constraints on the jet speed and inclinations of the two corresponding jet component pairs detected in EVN epoch 2. Shaded regions around the curves show the $1\sigma$ uncertainties. The green shaded region indicates the inclination angle of GRS~1915+105 measured prior to the obscured phase \citep{Reid2023}.
}
\label{fig:beta_theta}
\end{figure}

Assuming the approaching and receding jets are intrinsically symmetric and were launched simultaneously, we can use the relative distances from the core of corresponding jet components in EVN~E2 to calculate possible combinations of the jet speed $\beta$ and inclination angle $i$ \citep[e.g.,][]{Mirabel1999}:
\begin{equation}
    \frac{\Delta r_\mathrm{a}}{\Delta r_\mathrm{r}} = \frac{1 + \beta\cos{i}}{1 - \beta\cos{i}}
    .
\end{equation}
Here $\Delta r_\mathrm{a}$ and $\Delta r_\mathrm{r}$ are the angular separations from the core of the approaching and receding jet components, respectively, $\beta=v/c$ is the jet velocity, and $i$ is the inclination angle.
For the jet component pair closest to the core we find $\Delta r_\mathrm{a}/\Delta r_\mathrm{r}=1.28 \pm 0.06$ while for the second, more distant pair we find $\Delta r_\mathrm{a}/\Delta r_\mathrm{r}=1.25 \pm 0.02$. 
In Figure \ref{fig:beta_theta} we show the allowed combinations of $\beta$ and $i$ for the jet component pairs (contours from Fig.~\ref{fig:VLBA_threeplots}c). Error margins are propagated considering both statistical errors and systematic errors, following \cite{Pradel2006} and \citet{Reid_2017}. The constraints from both jet component pairs overlap within error margins.

We show the pre-obscuration inclination angle of $64 \pm 4^\circ$ \citep{Reid2023} as a shaded region in Figure~\ref{fig:beta_theta}. More recent estimates during the obscured phase find higher inclinations: $i=87 \pm 3^\circ$ \citep{Rodriguez2025}, who assumed the jet velocity remained constant at $\beta=0.80 \pm 0.12$, based on the consistency of proper motions measured over many years prior to the obscured phase \citep{Reid2023}. \citet{Jiang_2026} monitored the positions of discrete jet ejecta launched during the obscured phase over multiple days, and were able to directly constrain the velocity of these discrete ejecta to be $\beta \sim 0.35$, consequently constraining the inclination angle to $i=89 \pm 1^\circ$. This lower velocity measured by \cite{Jiang_2026} is inconsistent with the constant-velocity assumption of \cite{Rodriguez2025}; however, the authors note that their inclination estimate is largely insensitive to the assumed velocity. 
More recent observations with the EAVN from 2025 instead constrain the viewing angle of the jets to a smaller viewing angle, $i \lesssim 78^\circ$, and an intrinsic jet speed of $\beta \lesssim 0.37$ \citep{Yan_2026_2}, broadly similar to the velocity found by \citet{Jiang_2026}.

Our own constraints do not reach the high inclinations found by \citet{Rodriguez2025} and \citet{Jiang_2026} even for $\beta=1$, suggesting lower inclination angles during EVN E2. If we instead consider a jet velocity of $\beta \sim 0.35$, similar to that found by \citet{Jiang_2026} and consistent with the constraint from \citet{Yan_2026_2}, our constraints give inclination angles broadly consistent with the pre-obscuration estimate of $64 \pm 4^\circ$ \citep{Reid2023}, and with the upper limit of $i \lesssim 78^\circ$ found by \citet{Yan_2026_2}. However, we note that these estimates come from different epochs. If the position angle changes reported here indeed reflect jet precession, the inclination angle would be expected to vary between them as well, so close numerical agreement between epochs should not be taken as evidence for a single, unchanging inclination.

\begin{figure}
\centering
\includegraphics[width=1.0\columnwidth]{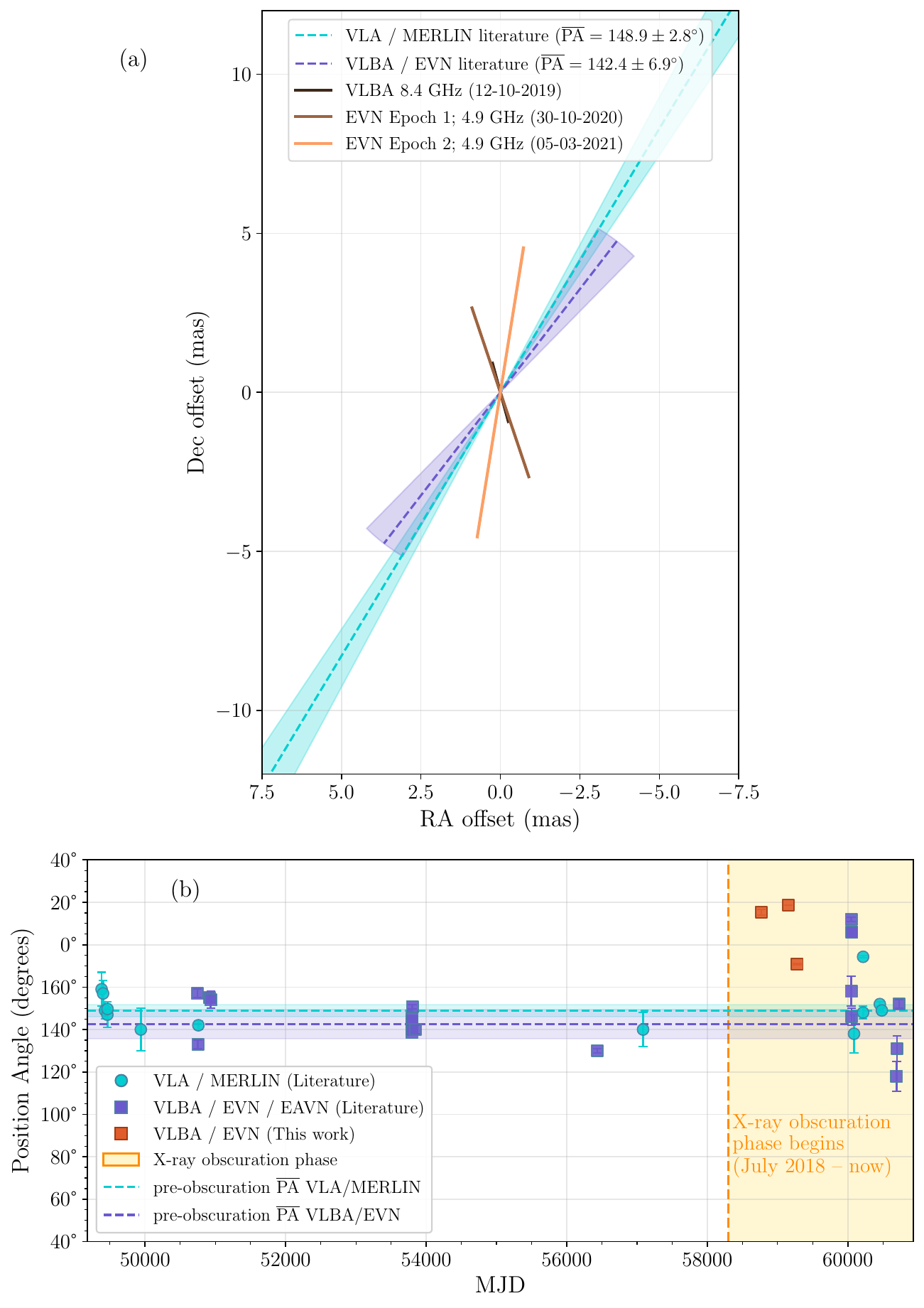}
\caption{\footnotesize
(a) Jet position angles measured with the VLBA and EVN in this paper (2019-2021), compared with the means of archival VLBA, EVN, EAVN, VLA and MERLIN measurements. The line lengths of the archival data are not to scale. (b) Evolution of the measured position angles as a function of time (MJD). The position angle axis is folded with a $180^\circ$ period, centred at $150^\circ$. The shaded region marks the X-ray obscuration phase beginning in July 2018 and continuing to the present. Individual position angle measurements and references are displayed in table~\ref{tab:pa_history}.
}
\label{fig:position_angles}
\end{figure}

\subsection{Variation in jet orientation}
\label{subsec:jet_changes}

The position angle of the jets from GRS~1915+105 have been detected in different accretion states over decades \citep{Rodriguez1999, Fender1999, Dhawan2000, Rushton_2007, Miller-Jones_2007, Reid_2014}. Before the obscured phase, the position angles measured on arcsecond and milliarcsecond scales span a range of $\sim $$30^\circ$ (from $130^\circ$ to $159^\circ$; see Fig. \ref{fig:position_angles} and Tab. \ref{tab:pa_history}), with a weighted mean of $144^\circ \pm 7^\circ$, though the reduced $\chi^2_\nu = 25$ suggests the formal uncertainties underestimate the true scatter, or that genuine intrinsic variability was present. Restricting to milliarcsecond-scale measurements alone, the pre-obscuration weighted mean is $142^\circ \pm 7^\circ$, which we adopt as the reference axis for comparison with our own milliarcsecond-scale observations below.

Our VLBI campaign, which began after the onset of the obscuration phase, shows a drastically different PA from the historical axis. In the first milliarcsecond-scale image of the
GRS~1915+105 jets after 2015, our VLBA data show a PA of $15.3^\circ \pm 1.2^\circ$. Our second milliarcsecond-scale image, taken with the EVN one year later in October 2020 (EVN~E1), shows a similar PA of $18.61^\circ \pm 0.03^\circ$, confirming that this large PA offset persists and is connected to the obscuration phase rather than being a one-off measurement. Our final detection, over four months after EVN~E1, shows the PA at $170.9^\circ \pm 0.3^\circ$ (EVN~E2), $28.9^\circ \pm 7.0^\circ$ from the pre-obscuration jet axis. These three detections fill an important gap in the jet orientation history between the onset of the obscuration phase in 2018 and the published milliarcsecond-scale position angles from 2023 \citep{Jiang_2026}, who subsequently measured a PA of $12^\circ$ with the EAVN, a value not seen at milliarcsecond scales prior to the obscured phase. This suggests that the jet's pointing direction is continuously changing throughout this phase.

The jet PA changes from $18.6^\circ$ (EVN E1) to $170.9^\circ$ (EVN E2) over 126 days, with a minimum angular displacement of $27.7^\circ$, implying a sky-projected slew rate $\dot{\rm PA} \gtrsim 0.2^\circ\,{\rm d}^{-1}$. The sparse temporal sampling prevents distinguishing whether the jet is steadily precessing or changing angle erratically. The position angles remain stable to within the fitting uncertainties ($\lesssim 5^\circ$) over the 8-hour EVN E1 observation (see Table \ref{tab:flux_positions}), implying a precession rate $\lesssim 5^\circ / 8\,{\rm hr} = 15^\circ {\rm d^{-1}}$, assuming the position angle changes are due to stable precession. We discuss the possible mechanisms responsible for the changes in jet position angle in Section~\ref{discussion} below.


\section{Discussion}\label{discussion}

Our VLBI epochs reveal jet position angle (PA) variations up to $57^{\circ} \pm 8^{\circ}$ with respect to the weighted mean of VLBI observations before the X-ray obscured phase (see Section \ref{subsec:jet_changes}). Arcsecond-scale VLA observations in 2023 also reported a change in jet PA, $\Delta$PA$=24^\circ$ \citep{Rodriguez2025}. This measurement traces discrete ejecta at much larger distances from the core, whose position reflects the jet's direction integrated over a long propagation time since launch. Our milliarcsecond-resolution VLBI observations, by contrast, probe the jet both spatially and temporally much closer to the point of launch, offering a more direct measurement of the jet-launching direction.

Placing this offset in the broader context of jet reorientation in X-ray binaries, the largest reported position-angle range is $\sim$110$^\circ$ for the neutron-star XRB Circinus X-1 \citep{Cowie2025}. This measurement combines observations from different instruments and resolutions spanning 25 years, and reflects the total peak-to-peak wandering of the axis in both directions over that baseline. Our 57$^\circ$ offset, by contrast, is a deviation in a single direction from a fixed pre-obscuration axis, measured consistently at milliarcsecond resolution, and represents the largest such offset reported from milliarcsecond-scale VLBI observations of an XRB. Milliarcsecond-scale EAVN observations in 2025 have since found the jet on the opposite side of this axis, at a PA of $118^\circ \pm 7^\circ$ \citep{Yan_2026_1}. Together, these measurements extend the range of jet PAs observed in GRS~1915+105 during the obscured phase to $\sim$80$^\circ$, which is the largest range reported for a black hole XRB.

Due to sparse temporal sampling we are unable to identify whether the jets in GRS~1915+105 are steadily precessing or changing direction erratically. EAVN observations from 2023 show a $6^\circ$ change within two days at $6.7$\,GHz ($\dot{\rm PA} \sim 3^\circ\,{\rm d}^{-1}$), and a $40^\circ$ difference between observations at $6.7$\,GHz and $43$\,GHz separated by two days ($\dot{\rm PA} \sim 20^\circ\,{\rm d}^{-1}$) \citep{Jiang_2026}. The latter comparison, however, spans two vastly different observing frequencies, so some or all of this difference could reflect a frequency-dependent core shift along a bent or precessing jet rather than a true fast change in jet direction over that interval. Since higher-frequency emission becomes optically thin closer to the black hole, the two frequencies probe different distances along the jet. If the jet is already curved due to ongoing precession, this alone could produce different, earlier-imprinted position angles at each frequency, without requiring the launching direction itself to have swung by $20^\circ$ within the two-day interval.

A genuine $20^\circ\,{\rm d}^{-1}$ position-angle change would require the reorientation rate to have increased substantially beyond the $\lesssim15^\circ\,{\rm d}^{-1}$ upper limit implied by the stability of our EVN E1 position angles over 8 hours (Section~\ref{subsec:jet_changes}). This either points to a core-shift origin for at least part of the $40^\circ$ difference, or, if the $40^\circ$ difference is genuine, indicates that the reorientation rate had already increased considerably beyond our E1-epoch constraint by 2023. The same-frequency $6.7$\,GHz rate of $\sim3^\circ\,{\rm d}^{-1}$, by contrast, is consistent with both of our EVN-derived limits ($\lesssim15^\circ\,{\rm d}^{-1}$ and $\gtrsim0.2^\circ\,{\rm d}^{-1}$). The temporal sampling remains too sparse, however, to establish whether this reflects a continuous precession, whether steady or accelerating, or an erratic reorientation, whether through fluctuating rates or discontinuous jumps. We consider possible physical origins for both below.

\subsection{Lense-Thirring precession}
\label{subsec: LT}

GRS~1915+105's transition into an X-ray obscured state raises the possibility of a mechanism analogous to that invoked for the 2015 outburst of the BHXB V404 Cygni, in which a
super-Eddington, X-ray obscured accretion episode was accompanied by rapid jet precession \citep{Miller-Jones_2019}. \cite{Motta2017} explain the X-ray obscuration using a slim disc model \citep{Abramowicz1998}, in which erratic or continuous super-Eddington accretion inflates the inner region of the accretion flow, hiding the innermost (and brightest) regions of the flow.

A slim disc scenario naturally gives rise to Lense–Thirring precession. Precession of the inner slim disc may drive jet precession, either through the magnetic field lines anchored in the disc, realignment of spin-powered jets via powerful inner disc outflows, or through direct influence of the precessing slim disc itself. 
For V404 Cygni's 2015 outburst, \cite{Miller-Jones_2019} estimated a Lense-Thirring precession timescale of $56 \, \rm s$.

The expected period of Lense-Thirring precession in the first order approximation is \citep{Fragile_2007, Middleton_2017}

\begin{equation}
    P_{prec} = \frac{\pi}{3 a_*} \frac{G M}{c^3} r^3_{sph} \left[  \frac{1-(r_{in}/r_{sph})^3}{ln(r_{sph}/r_{in})} \right]
\end{equation}
where $M$ is the mass of the central object; $a_*$ is the dimensionless black-hole spin $Jc/GM^2$ (with $J$ the spin angular momentum); $G$ is the gravitational constant; and $r_{in}$ and $r_{sph}$ are the inner and outer radii (in this case the spherization radius) of the slim disc, with all radii given in units of the gravitational radius $r_g = GM/c^2$. We assume a Kerr black hole with a mass of $12.4^{+2.0}_{-1.8} \, \rm M_{\odot}$ \citep{Reid_2014}, spin parameter $a_* = 0.703 \pm 0.034$ and $r_{in} = r_{ISCO} = 3.36 \pm 0.15 \, r_g$ \citep{Motta_2024}.

\cite{Miller-Jones_2019} estimated a spherization radius of tens to hundreds of $r_g$ for moderate spins for V404~Cyg, which is consistent with the maximum radius expected for rigid precession \citep{Motta_2017b}. For GRS~1915+105, a spherization radius equal to $r_{sph} = 100 \, r_g$ with these assumptions would result in a precession period of $P_{prec} \sim 27 \, \rm s$. A full derivation of $r_{\rm sph}$ following \citet{Poutanen_2007}, using GRS~1915+105's peak luminosity, gives a comparable or smaller spherization radius (of order $10\,r_g$), implying an even shorter precession timescale.

A precession period of this order cannot explain the observed timescales of the position angle changes reported here, which instead suggest that if precession is happening, it must involve a much larger portion of the accretion flow reaching the outer disc. Lense-Thirring precession also requires a slim-disc geometry, which recent XRISM Resolve spectroscopy disfavours, instead supporting a standard, sub-Eddington disc (Section~\ref{subsec:warping}; \citealt{Miller_2025}). The combination of an implausibly short predicted precession timescale and a disc geometry disfavoured by the best available spectroscopic data leads us to rule out Lense-Thirring precession as the primary driver of the observed jet reorientation.


\subsection{Radiation-driven warping}
\label{subsec:warping}

As noted above (Section~\ref{subsec: LT}), \citet{Miller_2025} model the first XRISM Resolve spectrum of GRS~1915+105 with a standard, sub-Eddington accretion disc, disfavouring the super-Eddington, slim-disc geometry that had remained viable in lower-sensitivity Chandra data. This disc temperature, together with the properties of the obscuring gas and the radii and inclinations inferred from the neutral FeK$\alpha$ and broad and narrow ionized emission lines, points to occultation of the central engine by the outer accretion disc itself. The line spectrum is further fit by three photoionized emission zones with small but significantly different bulk velocities, which the authors suggest may indicate axial asymmetry from a large-scale warp in the outer disc. Radiation pressure on the outer disc provides a viable mechanism for such a warp \citep{Pringle_1996} and the system's high inclination angle makes it plausible that a warped outer disc could occult our view of the central engine.

Spectral analysis of AstroSat, NICER and NuStar data constrains the distance of the absorbing plasma to within $\leq 2 \times 10^{10} \, \rm cm$ from the black hole \citep{Athulya_2023}. Earlier, \citet{Neilsen_2020} found an absorbing distance of $\sim few \times 10^{11} \, \rm cm$ from the black hole based on NICER data, arguing for a vertically extended stratified disc. Assuming a black hole mass of $12.4^{+2.0}_{-1.8} \, \rm M_{\odot}$ \citep{Reid_2014}, a distance of $10^{10}-10^{11} \, \rm cm$ corresponds to $\sim$$10^3-10^4 \, \rm R_g$, placing it in the mid-disc region where slim discs and radiation-driven warps could both operate \citep{Abramowicz1998,Ogilvie_2001}.

Radiation-driven warped discs can precess on super-orbital timescales \citep[e.g.][]{Maloney_1997}. If the warp-propagation timescale is shorter than the precession timescale the warp can propagate across the entire disc and precess as a global structure \citep{Fragile_2007, Liska_2018}. As the disc and hence the jet launching region precess, this is expected to cause the jet to precess. Strong evidence for a radiation-driven warped, precessing disc in an X-ray binary was provided by the discovery of the relativistic precessing jets in SS~433 \citep{Margon_1984}. The systematic velocity variations in the optical jet emission, the radio jet morphology, and the optical photometry of the system all indicate a disc precession period of 164 days. 

\citet{Maloney_1997} originally interpreted this as evidence for a global mode of precession, in which both the inner disc (explaining the jets) and the outer disc (explaining the photometry) precess together at the same rate. More recent work \citep{Begelman2006}, however, proposes a picture in which the jets remain fixed to the black hole spin axis and are instead bent into alignment with a precessing outflow launched from the warped outer disc, with the jets and photometric period coinciding for this reason rather than through direct co-precession of the jet-launching region. This mechanism specifically requires hyper-Eddington accretion, since the outflow is radiatively driven from well outside the black hole, and is therefore only expected to operate in this form during correspondingly high accretion states. Such a state is disfavoured for GRS~1915+105 by recent XRISM Resolve spectroscopy, which instead supports a standard, sub-Eddington disc \citep{Miller_2025}.

In order for a radiation-driven instability to arise, the irradiating luminosity must exceed a critical fraction of the Eddington luminosity such that the radiation reaction torque overcomes viscous damping.
Radiation-driven instability is strongest at larger radii where the disc intercepts more flux per unit surface mass, so the disc must extend to radii where the instability criterion is satisfied before being tidally truncated \citep{Wijers_1999}. Only those X‑ray binaries with sufficiently small mass ratios ($q = M_2/M_1$ ; with $M_1$ the mass of the compact object and $M_2$ the mass of the companion star) and sufficiently large orbital separations fall into the region of parameter space where persistent, steadily precessing radiation‑driven warps are expected \citep{Ogilvie_2001}.

This makes GRS~1915+105 a strong candidate for radiation-driven warping: with an orbital period of 33.9 days \citep{Steeghs_2013}, the longest of all LMXBs, its large orbital separation ensures the disc extends well into the instability regime. Furthermore, its small mass ratio ($q \sim 0.07$; using $M_1 = 12.4^{+2.0}_{-1.8} \, \rm M_{\odot}$ \citep{Reid_2014} and $M_2 = 0.81 \pm 0.53 \, \rm M_{\odot}$ \citep{Harlaftis_2004}) means the disc is only weakly tidally truncated, allowing it to extend to the large radii where the radiation-driven instability is strongest.

However, only a narrow range of orbital separations ($r_b$) can result in a steadily precessing, radiation-driven warped disc. Assuming a viscous parameter value of $\alpha = 0.3$ and accretion efficiency $\epsilon = L_*/{\dot M} c^2 = 0.1$, \cite{Ogilvie_2001} determined the ranges of stability (defined as containing a single unstable mode) to be $1.82 \leq r_b/( 10^6 \times G M_1 /c^2 ) \leq 2.79$ and $1.82 \leq r_b/( 10^6 \times G M_1 /c^2 ) \leq 2.63$ for a constant and variable luminosity respectively. Using the best known donor star mass estimate of $M_d = 0.81 \pm 0.53 \, \rm M_{\odot}$ for GRS~1915+105 \citep{Harlaftis_2004} and $\alpha = 0.3$ and $\epsilon = 0.1$, we get an $r_b$ that lies slightly above the upper bound of the region of stability ($r_b \sim 4 \times 10^6 \, G M_1 /c^2 $). This implies the existence of multiple interacting warping modes. 

The exact range of stability is uncertain, as it depends sensitively on the assumed values of $\alpha$ and $\epsilon$. Lower values would shift the instability threshold to higher $r_b$, potentially placing GRS~1915+105 within the region of stable, single-mode warping. The donor star mass estimate, and therefore $r_b$ itself, also carries significant uncertainty.

A genuine reduction in $\alpha$ and/or $\epsilon$ around this time is independently supported by observations: \citet{Motta2021} report that GRS~1915+105 entered a canonical low/hard accretion state in 2018, immediately preceding the onset of the X-ray obscured phase. This transition is characteristically associated with a less radiatively efficient, geometrically thicker inner accretion flow \citep[e.g.][]{Esin1997}. The source's continued decline toward its lowest recorded activity level in 2026 \citep{Motta2026atel, Marino2026atel, Ducoin2026atel,
CotiZelati2026atel} is consistent with this same trend continuing. Given GRS~1915+105's proximity to the region of stable, single-mode warping, a genuine hard-state transition provides a plausible trigger for the emergence of a stable warp. The specific character of the warp depends on the system's mass ratio and the location of mass input to the disc, as discussed below.

In the linear stability analysis of \citet{Ogilvie_2001}, radiation-driven disc warping is described in terms of discrete eigenmodes of the disc's bending-wave equation, each characterized by its radial structure (the number of nodes at which the disc's tilt reverses sign) and an associated growth rate and precession frequency. The lowest-order mode, mode 0, has no radial node: the disc tilts coherently in a single direction at all radii, and precesses retrogradely. This is the mode thought to operate in systems such as Her~X-1 and SS~433. Mode 1, the next-lowest-order mode, has one radial node, so that the inner and outer disc tilt in opposite directions, and precesses prograde.

Which of these modes is unstable first depends on the mass ratio $q$ and on where mass is added to the disc. For mass added at the circularization radius, \citet{Ogilvie_2001} find that mode 1 becomes unstable before mode 0 for $q\lesssim0.2$. If instead mass is added at the disc's outer radius, which is arguably more realistic given that the accretion stream is expected to impact the outer disc, mode 0 requires a much larger disc to become unstable at all, and mode 1 dominates over a substantially wider range of parameters. \citet{Ogilvie_2001} caution, however, that this result neglects the angular momentum deficit of mass added away from a circular orbit at $r_\mathrm{o}$, which would introduce an extra torque that might affect the stability properties.

With $q\sim0.07$ for GRS~1915+105, the system lies in the regime where mode 1 is expected to dominate under either assumption. Given the system's high inclination, mode 1's one-node structure could naturally place the inner and outer disc on opposite sides of the line of sight, potentially obscuring the inner disc regardless of the precession phase. This could naturally explain the persistent nature of the obscuration.

We also consider the scenario in which GRS~1915+105's accretion disc contains multi-modal warping. This would result in quasi-periodic and potentially less coherent precession compared to systems such as SS~433 and Her~X-1. 
In the linear regime, this can be represented by a superposition of global bending normal waves, each with their own growth rate and precession frequencies. However, once the warp amplitude becomes large, non-linear effects modify the internal torque, which can reach a critical amplitude beyond which the steepest part of the warp steepens further, potentially leading to disc breaking \citep{Ogilvie_1999, Ogilvie_2000, Dogan_2018}. Additionally, parametric instability can develop, in which inertial waves extract energy from the warp, driving it toward decay or a turbulent regime \citep{Gammie_2000, Fragile_2026}.

If GRS~1915+105 sits in the multi-modal regime, the observed jet reorientation may reflect non-linear, irregular warping rather than a clean, steadily precessing global mode. This is broadly consistent with the increased rate of position angle change inferred from the 2023 EAVN campaign, though our data cannot distinguish this from an accelerating continuous precession or a frequency-dependent effect (Section~\ref{discussion}). Multi-modal warping is, however, harder to reconcile with the persistent, multi-year nature of the obscuration observed since 2019, which would more naturally suggest a stable, long-lived disc structure.

Independent evidence for a warped disc in GRS~1915+105 itself predates the current obscured phase. \citet{Rau2003} reported a $590$-day periodicity in the source's hard X-ray and radio flux from over four years of RXTE monitoring, and put forth precession of a radiation-driven warp as an explanation, by analogy with the super-orbital periods seen in other warped-disc systems. This periodicity, however, corresponds to only a few cycles, and could plausibly instead reflect red noise. Stochastic, intrinsic source variability could mimic a periodic signal when only a small number of cycles are sampled \citep{Vaughan2005}. The modulation was notably absent from the soft, disc-dominated X-ray flux, unlike Her~X-1, SMC~X-1, and LMC~X-4, where the soft X-ray flux carries the super-orbital signal. The authors therefore only tentatively identified the periodicity's physical origin with disc warping. If a fixed-period, precessing warp was indeed already present prior to 2018, the hard-state transition discussed above would not have established single-mode warping for the first time. Instead, it would have driven an increase in the warp's amplitude, growing it from a level insufficient to occult the central engine into the large-amplitude warp required to produce the persistent obscuration and large PA variations.

Radiative warping therefore offers a plausible mechanism for the observed jet precession and obscuration in GRS~1915+105. The system's proximity to the single-mode stability boundary, combined with the plausible reduction in $\alpha$ and $\epsilon$ associated with the source's hard-state transition and subsequent decline, makes a stable, single-mode warp a viable candidate, and one that naturally accounts for the persistent nature of the obscuration. However, we cannot exclude a more irregular, multi-modal warp, and GRS~1915+105's proximity to the stability boundary may equally place it in a regime that is not a clean single-mode. Furthermore, mode~1 radiative warping has not been confirmed in any X-ray binary, and dedicated simulations in this regime have yet to be performed.

\subsection{Other explanations}
\label{subsec: other expl}
Other processes have been proposed to explain the apparently sudden onset of jet precession in GRS~1915+105. Previous studies have invoked interactions with an external perturber \citep{Rodriguez2025, Jiang_2026}. 
Recent population studies suggest that triple evolution is a significant formation channel for BH-LMXBs \citep{Shariat_2025}. However, the only current evidence of a BH-LMXB in a triple system is V404 Cyg whose tertiary companion is inferred to be at least 3500 astronomical units (AU) away \citep{Burdge_2024}. Simulations find that classical triple-channel formation without a common envelope typically results in outer separations of $10^3-10^4 \, \rm AU$, while a common envelope phase produces less constrained separations, generally in the range  $\sim$$100-1000 \, \rm AU$. 
However, wider inner separations can result in closer outer separations \citep{Shariat_2025}, potentially bringing a tertiary companion close enough to be dynamically relevant. Nevertheless, GRS~1915+105's decades-long X-ray brightness would require any tertiary companion to reside in a stable, wide orbit, making a dynamical perturbation sufficient to produce the observed precession onset difficult to motivate without additional evidence. 
Additionally, if the recent multi-wavelength non-detections \citep{Motta2026atel, Marino2026atel, Ducoin2026atel, CotiZelati2026atel} indeed indicate a genuine transition into a low-accretion state, rather than a further deepening of the obscuration, this would favour an intrinsic, accretion-driven origin for the jet reorientation over an external perturbation, since the latter would not obviously predict a large-scale drop in luminosity.

Other studies have suggested that the obscuration could be the result of disc winds enshrouding the central emitting region \citep{Miller_2020,Ghandi_2025}.
This explanation would be consistent with the predictions of \citet{Fender2026}, who found a correlation between the presence of absorption and the observations of slow and precessing jets. One possible physical link is mechanical wind-induced bending of the jets, which has been detected in the high mass X-ray binary Cygnus X-1 \citep{Prabu2026}. However, this is for winds expelled by the companion star, and simulations show that sufficiently strong stellar-wind bending is expected to disrupt the jet into a diffuse, no longer collimated outflow \citep{Yoon_2015}. \citet{Jiang_2026} invoke this stellar-wind interaction physics by analogy to argue against a disc-wind origin for GRS~1915+105, based on the largely ballistic motion of the observed discrete ejecta. Disc-wind-driven jet bending, distinct from the stellar-wind case, has also been proposed directly for the precessing microquasar SS~433, where a powerful wind driven off a hyper-Eddington outer disc bends otherwise spin-aligned jets into alignment with the disc's precession \citep{Begelman2006}, as discussed in Section~\ref{subsec:warping}. However, this mechanism specifically requires hyper-Eddington accretion, in tension with the sub-Eddington, standard-disc geometry favoured by recent XRISM spectroscopy of GRS~1915+105 (Section~\ref{subsec: LT}; \citealt{Miller_2025}), and its applicability to the current, lower-accretion-rate obscured phase therefore remains uncertain.

\section{Conclusions}\label{Conclusions}

We observed GRS~1915+105 for four epochs using VLBI during its X-ray obscured state between 2019--2021. As the earliest and highest-resolution radio observations of GRS~1915+105 during this phase, these data resolve the jet on milliarcsecond scales, closer to launch where accretion-outflow coupling is stronger.

The jet position angle evolved from $15.3^{\circ}\pm1.2^{\circ}$ (VLBA) to
$18.61^{\circ}\pm0.03^{\circ}$ (EVN~E1), before shifting by a minimum of $27.7^{\circ}$ to $170.9^{\circ}\pm0.3^{\circ}$ (EVN~E2) over the following 126 days with no source detected at EVN~E3. This sequence corresponds to a maximum offset of $57^{\circ}\pm8^{\circ}$ from the weighted mean of jet angles measured prior to the X-ray obscured phase. We find no evidence for jet orientation changes within our 8-hour EVN~E1 observation, implying a precession rate $\lesssim 15^\circ \rm d^{-1}$ if the changes are due to stable precession.

We rule out Lense-Thirring precession of the inner slim disc as the direct driver. The predicted precession timescale is orders of magnitude shorter than observed, and the slim-disc geometry it requires is disfavoured by recent XRISM Resolve spectroscopy, which instead supports a standard, sub-Eddington disc.

We instead find that radiation-driven disc warping, previously proposed to explain the
persistent X-ray obscuration of GRS~1915+105 \citep{Miller_2025}, also plausibly accounts for the observed jet reorientation and its apparent coincidence with the onset of obscuration. 
GRS~1915+105 lies close to the boundary of the regime in which a single, steadily precessing mode~1 warp is expected, which would naturally explain the persistent, multi-year obscuration. We suggest that the source's transition into a low/hard accretion state, and its continued decline toward its lowest recorded activity level in 2026, may have driven the emergence or growth of such a warp. However, the parameters placing GRS~1915+105 near this boundary carry significant uncertainties and mode~1 warping has not been confirmed in any X-ray binary. 
Sparse temporal sampling also prevents us from establishing whether the jet reorientation is smooth or erratic, so we cannot exclude a warp in which multiple modes, precessing at different rates, are superposed rather than one coherent structure dominating.


\begin{acknowledgements}
LIR would like to thank Benito Marcote for useful discussions on VLBI imaging.

PA and LIR are supported by the WISE fellowship program, which is financed by NWO.
SEM acknowledges support from the INAF Fundamental Research Grant (2022) EJECTA.
RPF acknowledges support from the UKRI, European Research Council Synergy Grant 'Blackholistic' (grant agreement number 101071643) and The Hintze Family Charitable Foundation.

The European VLBI Network is a joint facility of independent European, African, Asian, and North American radio astronomy institutes. Scientific results from data presented in this publication are derived from the following EVN project code: EM144. We acknowledge support from the Astrophysics Center for Multi-messenger Studies in Europe (ACME), funded under the European Union's Horizon Europe Research and Innovation Program, Grant Agreement No. 101131928.

The National Radio Astronomy Observatory is a facility of the National Science Foundation operated under cooperative agreement by Associated Universities, Inc. This work made use of the Swinburne University
of Technology software correlator, developed as part of
the Australian Major National Research Facilities Programme and operated under licence.

We thank the staﬀ of the Mullard Radio Astronomy Observatory,
University of Cambridge, for their support in the maintenance and
operation of AMI. We thank the staﬀ at the South African Radio
Astronomy Observatory (SARAO) for scheduling the MeerKAT
observations. The MeerKAT telescope is operated by the South
African Radio Astronomy Observatory, which is a facility of the
National Research Foundation, an agency of the Department of
Science and Innovation.
This research has made use of the MAXI data provided by RIKEN, JAXA and the MAXI team \citep{Matsuoka_2009}.

The authors wish to recognize and acknowledge the
very significant cultural role and reverence that the summit of Maunakea has always had within the indigenous
Hawaiian community. We are most fortunate to have the
opportunity to conduct observations from this mountain.

\end{acknowledgements}

\section{Data Availability}\label{Section 8}

The uncalibrated VLBA visibilities presented in this paper are publicly available in the National Radio Astronomy Observatory archive under project code BM500. The uncalibrated visibilities of the three EVN epochs are publicly available in the EVN archive under project code EM144 (\url{https://doi.org/10.48717/y0rm-hd58}). The uncalibrated MeerKAT visibility data are publicly available in the South African Radio Astronomy Observatory archive at \url{https://archive.sarao.ac.za}, subject to a standard proprietary period of one year. The AMI-LA data are available upon reasonable request to the corresponding author. Data from Swift/BAT are publicly available in NASA's HEASARC archive (\url{https://heasarc.gsfc.nasa.gov}). The MAXI/GSC data are provided by RIKEN, JAXA, and the MAXI team (\url{http://maxi.riken.jp}). Data not available through public archives, and all source code, will be shared upon reasonable request to the corresponding author.

\bibliographystyle{aa} 
\bibliography{bibliography} 

\begin{appendix}

\onecolumn

\section{Position angle compilation}

\begin{table}[ht]
\centering
\caption{Compiled position angle (PA) measurements of the GRS~1915+105 jet. PA values are measured east of north and are defined modulo $180^\circ$, reflecting the undirected orientation of the fitted jet axis. Where both approaching and receding components were detected, the PA represents the average of the two.}
\label{tab:pa_history}
\begin{tabular}{llclcl}
\hline\hline
Epoch & MJD & PA ($^\circ$) & Instrument & $\nu_c$ (GHz) & References \\
\hline
1994 Jan 29 & 49381 & $159 \pm 8$ & VLA  & 8.4  & \cite{Rodriguez1999} \\
1994 Feb 19 & 49402 & $157 \pm 6$ & VLA  & 8.4 & \cite{Rodriguez1999} \\
1994 Mar 19 & 49430 & $149 \pm 4$ & VLA  & 8.4 & \cite{Rodriguez1999} \\
1994 Apr 21 & 49463 & $147 \pm 6$ & VLA  & 8.4 & \cite{Rodriguez1999} \\
1994 Apr 23 & 49465 & $149.7 \pm 0.7$ & VLA  & 8.4 & \cite{Rodriguez1999} \\
1995 Aug 10 & 49939 & $140 \pm 10$ & VLA  & 8.4 & \cite{Rodriguez1999} \\
1997 Oct 23 & 50744 & $157 \pm 2$ & VLBA  & 15 & \cite{Dhawan2000} \\
1997 Oct 31 & 50752 & $133 \pm 2$ & VLBA  & 2.3 & \cite{Dhawan2000} \\
1997 Nov 6  & 50758 & $142 \pm 2$ & MERLIN  & 5.0  & \cite{Fender1999} \\
1998 Apr 11 & 50914 & $155 \pm 2$ & VLBA & 15 & \cite{Dhawan2000} \\
1998 May 2  & 50935 & $154 \pm 4$ & VLBA  & 15 & \cite{Dhawan2000} \\
2006 Feb 8  & 53794 & $144.1 \pm 1$ & VLBA & 8.4 & \cite{Miller-Jones_2007} \\
2006 Feb 8  & 53794 & $140.1 \pm 1$ & VLBA & 2.3 & \cite{Miller-Jones_2007} \\
2006 Feb 12 & 53798 & $143.4 \pm 1$ & VLBA & 8.4 & \cite{Miller-Jones_2007} \\
2006 Feb 14 & 53800 & $145.3 \pm 1$ & VLBA & 8.4 & \cite{Miller-Jones_2007} \\
2006 Feb 14 & 53800 & $138.5 \pm 1$ & VLBA & 2.3 & \cite{Miller-Jones_2007} \\
2006 Feb 17 & 53803 & $150.7 \pm 1$ & VLBA & 8.4 & \cite{Miller-Jones_2007} \\
2006 Apr 21 & 53846 & $140 \pm 2$ & eEVN & 5.0 & \cite{Rushton_2007} \\
2013 May 24 & 56436 & $130 \pm 1$ & VLBA  & 22  & \cite{Reid_2014} \\
2015 Mar 5  & 57086 & $140 \pm 8$ & VLA  & 22  & 15A-460 (PI: A. King) \\

2019 Oct 12 & 58768 & $15.3 \pm 1.2$ & VLBA  & 8.4  & \textbf{This paper} \\
2020 Oct 30 & 59152 & $18.61 \pm 0.03$ & EVN  & 4.9  & \textbf{This paper} \\
2021 Mar 5  & 59278 & $170.9 \pm 0.3$ & EVN  & 4.9  & \textbf{This paper} \\

2023 Apr 15 & 60049 & $12 \pm 1$ & EAVN & 6.7 & \cite{Jiang_2026} \\
2023 Apr 16 & 60050 & $158 \pm 7$ & EAVN & 43 & \cite{Jiang_2026} \\
2023 Apr 17 & 60051 & $6 \pm 1$ & EAVN & 6.7 & \cite{Jiang_2026} \\
2023 Apr 19 & 60053 & $146 \pm 4$ & EAVN & 43 & \cite{Jiang_2026} \\
2023 Apr 23 & 60057 & $6 \pm 1$ & EAVN & 6.7 & \cite{Jiang_2026} \\
2023 May 23 & 60088 & $138 \pm 9$ & VLA  & 5.5  & SS192168 (PI: A. Shaw) \\
2023 Sep 30 & 60217 & $148 \pm 3$ & VLA  & 10  & \cite{Jiang_2026} \\
2023 Sep 30 & 60217.5 & $174.3 \pm 0.5$ & VLA  & 8.0 & 23B-314 (PI: S.E. Motta) \\
2024 May 25 & 60455 & $152 \pm 1$ & VLA  & 5.5  & 24A-474 (PI: S.E. Motta) \\
2024 Jun 23 & 60484 & $149 \pm 1$ & VLA  & 5.5  & 24A-474 (PI: S.E. Motta) \\
2025 Jan 18 & 60693 & $118 \pm 7$ & EAVN & 6.7 & \cite{Yan_2026_1} \\
2025 Jan 25 & 60700 & $131 \pm 6$ & EAVN & 6.7 & \cite{Yan_2026_1} \\
2025 Feb 24 & 60730 & $152 \pm 2$ & EAVN & 6.7 & \cite{Yan_2026_1} \\

\hline
\end{tabular}

\end{table}

\newpage

\section{Time binning results from EVN E1}

\begin{table*}[ht]
\centering
\setlength{\tabcolsep}{4pt}
\caption{
Flux and position angle (PA) measurements of the time bins in EVN E1. Values correspond to Figure \ref{fig:EM144A}. The peak and total flux density measurements and position angle (PA) measurements are derived following self-calibration on 10 minute intervals, to improve the amplitude calibration.
}
\label{tab:flux_positions}
\begin{tabular}{c c c c}
\hline\hline
Time Range & Peak Flux Density & Total Flux Density & PA \\
(UTC) & (mJy\,beam$^{-1}$) & (mJy) & ($^\circ$) \\
\hline
\hline
12:30--13:00 & $6.23 \pm 0.09$    & $102.8 \pm 1.6$  & $20.9 \pm 2.0$  \\
13:00--13:30 & $6.16 \pm 0.10$    & $98.4 \pm 1.8$   & $21.3 \pm 2.1$  \\
13:30--14:00 & $5.50 \pm 0.09$    & $93.8 \pm 1.7$   & $19.9 \pm 2.1$  \\
14:00--14:30 & $6.28 \pm 0.14$    & $95.5 \pm 2.2$   & $19.1 \pm 2.7$  \\
14:30--15:00 & $11.67 \pm 0.21$   & $96.6 \pm 1.9$   & $18.8 \pm 2.8$  \\
15:00--15:30 & $7.86 \pm 0.16$    & $93.7 \pm 2.1$   & $18.3 \pm 2.6$  \\
15:30--16:00 & $9.04 \pm 0.20$    & $95.6 \pm 2.3$   & $20.4 \pm 2.9$  \\
16:00--16:30 & $10.31 \pm 0.22$   & $99.7 \pm 2.4$   & $19.9 \pm 3.0$  \\
16:30--17:00 & $11.28 \pm 0.19$   & $97.0 \pm 1.8$   & $20.6 \pm 2.5$  \\
17:00--17:30 & $13.07 \pm 0.45$   & $97.4 \pm 3.7$   & $21.6 \pm 7.9$  \\
17:30--17:40 & $15.91 \pm 0.20$   & $105.0 \pm 1.5$  & $22.3 \pm 2.5$  \\
17:40--17:50 & $16.94 \pm 0.19$   & $108.4 \pm 1.4$  & $22.0 \pm 2.4$  \\
17:50--18:00 & $17.85 \pm 0.24$   & $111.8 \pm 1.8$  & $21.7 \pm 2.9$  \\
18:00--18:05 & $21.32 \pm 0.31$   & $115.1 \pm 1.9$  & $24.5 \pm 3.0$  \\
18:05--18:10 & $20.01 \pm 0.23$   & $117.1 \pm 1.5$  & $22.2 \pm 2.5$  \\
18:10--18:15 & $22.13 \pm 0.26$   & $123.8 \pm 1.7$  & $21.4 \pm 2.6$  \\
18:15--18:20 & $23.73 \pm 0.30$   & $132.1 \pm 1.9$  & $21.1 \pm 3.0$  \\
18:20--18:25 & $25.09 \pm 0.31$   & $135.7 \pm 1.9$  & $22.1 \pm 3.1$  \\
18:25--18:30 & $26.53 \pm 0.30$   & $138.4 \pm 1.8$  & $20.7 \pm 2.9$  \\
18:30--18:35 & $27.44 \pm 0.27$   & $142.2 \pm 1.7$  & $21.4 \pm 2.5$  \\
18:35--18:40 & $28.49 \pm 0.33$   & $148.5 \pm 2.0$  & $21.0 \pm 3.1$  \\
18:40--18:45 & $30.90 \pm 0.33$   & $156.4 \pm 2.0$  & $21.0 \pm 3.0$  \\
18:45--18:50 & $32.15 \pm 0.30$   & $161.5 \pm 1.8$  & $21.4 \pm 2.6$  \\
18:55--19:00 & $36.06 \pm 0.32$   & $173.1 \pm 1.8$  & $21.8 \pm 2.6$  \\
19:00--19:05 & $36.87 \pm 0.30$   & $172.2 \pm 1.7$  & $21.7 \pm 2.5$  \\
19:05--19:10 & $38.55 \pm 0.27$   & $178.7 \pm 1.5$  & $22.1 \pm 2.2$  \\
19:10--19:15 & $42.63 \pm 0.32$   & $184.4 \pm 1.7$  & $20.4 \pm 2.4$  \\
19:15--19:20 & $45.09 \pm 0.36$   & $191.1 \pm 1.9$  & $20.1 \pm 2.6$  \\
19:20--19:25 & $47.16 \pm 0.42$   & $205.2 \pm 2.2$  & $23.0 \pm 3.2$  \\
19:25--19:30 & $66.81 \pm 0.74$   & $213.0 \pm 3.0$  & $25.4 \pm 4.4$  \\
19:30--19:35 & $112.56 \pm 0.59$  & $216.3 \pm 1.6$  & $20.8 \pm 1.5$  \\
19:35--19:40 & $93.63 \pm 3.06$   & $190.8 \pm 8.8$  & $9.1 \pm 15.7$  \\
19:40--19:45 & $138.81 \pm 0.96$  & $228.9 \pm 2.4$  & $16.9 \pm 4.0$  \\
19:45--19:50 & $143.61 \pm 0.99$  & $237.2 \pm 2.4$  & $17.5 \pm 3.8$  \\
19:50--19:55 & $144.89 \pm 0.74$  & $247.1 \pm 1.9$  & $17.6 \pm 3.1$  \\
19:55--20:00 & $146.29 \pm 0.70$  & $254.6 \pm 1.8$  & $16.5 \pm 3.0$  \\
20:00--20:05 & $148.20 \pm 0.76$  & $257.7 \pm 1.9$  & $14.4 \pm 3.2$  \\
20:05--20:10 & $157.00 \pm 0.61$  & $268.6 \pm 1.5$  & $15.9 \pm 2.6$  \\
20:10--20:15 & $158.92 \pm 0.63$  & $272.5 \pm 1.6$  & $15.5 \pm 2.8$  \\
\hline
\end{tabular}
\end{table*}

\end{appendix}
\end{document}